\documentclass[aps,pre,twocolumn,showkeys,groupedaddress,showkeys]{revtex4-2}
\usepackage{eurosym}
\usepackage{amsmath}
\usepackage{graphicx}
\usepackage{diagbox}
\usepackage{xcolor}
\usepackage{tabularx}
\usepackage[colorlinks=true, letterpaper=true, pdfstartview=FitV, linkcolor=blue, citecolor=blue, urlcolor=blue]{hyperref}

\begin{document}

\title{Quiescent and traveling solitons in the fractional parametrically
driven damped nonlinear Schr\"{o}dinger equation}
\author{Dongdong Wang$^{1,2}$}
\author{Rujiang Li$^{3}$}
\author{David Laroze$^{4}$}
\author{Boris A. Malomed$^{4,5}$}
\author{Pengfei Li$^{1,2}$}
\email{lpf281888@gmail.com}
\affiliation{$^{1}$Department of Physics, Taiyuan Normal University,
	Jinzhong 030619, China}
\affiliation{$^{2}$Institute of Computational and
	Applied Physics, Taiyuan Normal University, Jinzhong 030619, China}
\affiliation{$^{3}$National Key Laboratory of Radar Detection and Sensing,
	School of Electronic Engineering, Xidian University, Xi'an 710071, China}
\affiliation{$^{4}$Instituto de Alta Investigaci\'{o}n, Universidad de
	Tarapac\'{a}, Casilla 7D, Arica, Chile}
\affiliation{$^{5}$Department of Physical Electronics, School of Electrical
	Engineering, Faculty of Engineering, and Center for Light-Matter
	Interaction, Tel Aviv University, Tel Aviv 69978, Israel}

\begin{abstract}
We systematically investigate the existence, stability, and dynamics of
optical solitons in the framework of the one-dimensional nonlinear Schr\"{o}%
dinger equation with the Riesz-fractional diffraction operator, cubic
self-focusing, and linear loss, balanced by a linear parametric drive. The
model, which can be realized in a laser cavity, produces standing and moving
solitons, the latter ones existing below a critical velocity. One of the
soliton species is stable in a wide range of parameters, while others are
unstable. The fractional diffraction significantly alters the existence
conditions and stability thresholds of the solitons. Collision between
moving solitons are considered too. The results essentially expand the
variety of nonlinear modes in media with fractional diffraction.
\end{abstract}

\maketitle

\section{Introduction}

Starting from the introduction of fractional quantum mechanics \cite{49},
the interplay of fractional diffraction or dispersion with a trapping
potential and/or material nonlinearity is a subject of many theoretical
studies \cite{1}, that are often based on fractional nonlinear Schr\"{o}%
dinger equations (NLSEs) \cite{2,3,4}. In these equations, the usual
Laplacian, which represents the diffraction operator, is replaced by a
combination of fractional derivatives of the Riesz type \cite{47}, which are
actually nonlocal (pseudodifferential) operators, characterized by the
respective L\'{e}vy index (LI) $\alpha $ \cite{48}. In this context, much
interest was drawn to various species of solitons, predicted as solutions of
the fractional NLSEs \cite{26,27,28,29,30,31,32,33,34,35,36,37,38} (see also
reviews \cite{2,3}). Usually, these models are formulated in terms of
optics, and their first experimental realization was reported in fiber-laser
cavities, which emulate fractional group-velocity dispersion (GVD) \cite{25}%
, as originally proposed by Longhi \cite{50}. To this end, the light beam,
in the Fourier-decomposed form, passes a specially designed phase shifter
(actually, a hologram), which introduces a distributed phase shift emulating
what is expected from the action of the fractional GVD; eventually, the
decomposed beam is recombined and fed back into the optical cavity.
Recently, the creation of temporal solitons in laser cavities combining the
effective fractional GVD and Kerr nonlinearity has been reported too \cite%
{43,44}.

In conservative fractional-diffraction systems, solitons are maintained, as
usual, by the balance between the diffraction and self-focusing nonlinearity
\cite{2,3,4}. If the system also included loss and gain, it may give rise to
dissipative solitons, which exist if the balance between the diffraction and
nonlinearity is supplemented by the equilibrium between the gain and loss
\cite{5,6}.

While the most frequently used gain, compensating the background loss of the
medium, is provided by the intrinsic linear amplification of signals, such
as the lasing mechanism \cite{45}, an alternative type of the gain can be
applied by an external parametric drive \cite{46,11,12,13,14}. In optics, it
can be realized using the parametric interaction of the photonic signal with
the pump beam. The respective model, that readily gives rise to solitons,
amounts to the parametrically driven damped nonlinear Schr\"{o}dinger
equation (PDDNLSE). In addition to optics, it occurs in other physical
settings, such as water waves \cite{8,9,10} and magnetics \cite{15,16b,16c}.

One-dimensional PDDNLSE with the self-focusing or defocusing cubic
nonlinearity provides exact stationary solutions for bright \cite{16,17} or
dark \cite{18} solitons. It also supports traveling-soliton solutions \cite%
{19,20,20a,21}, some of which are stable \cite{19}. The driven-damped solitons
can build bound states in the form of steady or oscillating complexes \cite%
{22}. The bound states, along with the traveling solitons in the
conservative (undamped) limit, have been studied in detail, both
analytically and numerically, revealing sophisticated stability landscapes
and complex dynamical regimes \cite{23,24}. Stable dissipative solitons can
also be produced by complex Ginzburg-Landau equations with the parametric
gain \cite{51}.

The subject of the present work is the interplay of the parametric drive and
loss with fractional diffraction and intrinsic nonlinearity, which was not
studied before. This setting is modeled by the fractional version of
PDDNLSE, which is introduced below as Eq. (\ref{fmPDDNLSE}). The breaking of
the Galilean invariance by the driving term, the action of the loss and
parametric gain, and the nonlocal character of the fractional diffraction
give rise to many aspects of soliton phenomenology, including the existence
and stability of fractional dissipative solitons, the possibility of finding
moving fractional solitons, the formation and robustness of soliton bound
states, and the role of LI. The reported results offer guidance for new
experiments in fractional optical systems, as well as fluid dynamics and
other platforms which can implement the fractional wave dynamics.

The subsequent presentation is organized as follows. The model is introduced
in Section 2, which is followed by the systematic consideration of quiescent
solitons in Section 3. The analysis of moving solitons in the lossless model
is reported in Section 4, collisions between them being the subject of
Section 5. The paper is concluded by Section 6.

\section{The model}

The one-dimensional fractional PDDNLSE with the self-focusing cubic term,
for complex wave amplitude $\psi \left( x,t\right) $ of the optical field in
the laser cavity is written in the scaled form:

\begin{equation}
i\frac{\partial \psi }{\partial t}-iV\frac{\partial \psi }{\partial \xi }%
+(i\gamma +\omega )\psi -\left( -\frac{\partial ^{2}}{\partial \xi ^{2}}%
\right) ^{\alpha /2}\psi +2|\psi |^{2}\psi =h\psi ^{\ast }.
\label{fmPDDNLSE}
\end{equation}%
With the intention to include moving solitons below, Eq. (\ref{fmPDDNLSE})
is written with respect to the moving coordinate $\xi =x-Vt$, where $V$ is
the soliton's velocity (if any), $\gamma \geq 0$ is the loss coefficient,
real $\omega $ is the detuning of the parametric drive,$\ h>0$ is the
amplitude of the parametric drive (with $\ast $ standing for the complex
conjugate), and LI $\alpha $ defines the fractional Riesz derivative. If the
fractional NLSE is derived in the temporal domain for the fiber cavity, with
the fractional derivative representing GVD, $t$ and $x$ in Eq. (\ref%
{fmPDDNLSE}) denote the propagation distance and local time, respectively
\cite{25}.

As usual, we assume that LI belongs to interval $1<\alpha \leq 2$. Here,
values $\alpha \leq 1$ are irrelevant, as they give rise to the collapse of
the wave function \cite{2,3} , while $\alpha =2$ corresponds to the usual
(non-fractional) diffraction/dispersion operator. In the latter case, Eq. (%
\ref{fmPDDNLSE}) simplifies to \cite{19}

\begin{equation}
i\frac{\partial \psi }{\partial t}-iV\frac{\partial \psi }{\partial x}%
+(i\gamma +\omega )\psi +\frac{\partial ^{2}\psi }{\partial x^{2}}+2|\psi
|^{2}\psi =h\psi ^{\ast }.  \label{mPDDNLSE}
\end{equation}%
The standard form of PDDNLSE (\ref{mPDDNLSE}) pertains to quiescent
solitons, with $V=0$ and $\omega =-1$ fixed by rescaling \cite{16}:

\begin{equation}
i\frac{\partial \psi }{\partial t}+(i\gamma -1)\psi +\frac{\partial ^{2}\psi
}{\partial x^{2}}+2|\psi |^{2}\psi =h\psi ^{\ast }.  \label{PDDNLSE}
\end{equation}%
Equation (\ref{PDDNLSE}) admits two well-known exact stationary soliton
solutions,

\begin{equation}
\psi _{\pm }(x)=\exp \left( -i\theta _{\pm }\right) A_{\pm }\text{sech}%
\left( A_{\pm }x\right) ,  \label{Exact-sols}
\end{equation}%
with
\begin{gather}
\cos \left( 2\theta _{\pm }\right) =\pm \sqrt{1-\gamma ^{2}/h^{2}},
\label{COS} \\
A_{\pm }=\sqrt{1+h\cos \left( 2\theta _{\pm }\right) },  \label{Amp}
\end{gather}%
which exist under condition $\gamma ^{2}<h^{2}$. The known result \cite{15}
is that solutions $\psi _{-}$ (the one with the smaller amplitude) is always
unstable, while $\psi _{+}$ is stable for
\begin{equation}
h<\sqrt{1+\gamma ^{2}}.  \label{h<}
\end{equation}

In the absence of losses, $\gamma =0$, equation (\ref{PDDNLSE}) becomes a
conservative one:%
\begin{equation}
i\frac{\partial \psi }{\partial t}-\psi +\frac{\partial ^{2}\psi }{\partial
x^{2}}+2\left\vert \psi \right\vert ^{2}\psi =h\psi ^{\ast },
\label{conservativePDDNLSE}
\end{equation}%
with Lagrangian%
\begin{equation}
L=\text{Re}\int_{-\infty }^{+\infty }\left[ i\left( \psi ^{\ast }\psi
_{t}-\psi \psi _{t}^{\ast }\right) -|\psi |^{2}-\left\vert \psi
_{x}\right\vert ^{2}+|\psi |^{4}-h\psi ^{2}\right] dx.  \label{Lagrangian}
\end{equation}%
In this case, the complex exact soliton solutions, given by Eqs. (\ref%
{Exact-sols}), (\ref{COS}) and (\ref{Amp}), reduce to the real form:%
\begin{equation}
\psi _{\pm }(x)=\sqrt{1\pm h}\text{sech}\left( \sqrt{1\pm h}x\right) .
\label{conservative-sols1}
\end{equation}%
These solutions exist at $h<1$, the one $\psi _{+}$ being stable. It exists
also at $h>1$, being unstable in that case.

\section{Quiescent solitons}

To produce solutions of Eq. (\ref{fmPDDNLSE}) with $V=0$ for quiescent
solitons, we again fix $\omega =-1$ by means of scaling, rewriting the
underlying equation as

\begin{equation}
i\frac{\partial \psi }{\partial t}+(i\gamma -1)\psi -\left( -\frac{\partial
^{2}}{\partial x^{2}}\right) ^{\alpha /2}\psi +2|\psi |^{2}\psi =h\psi
^{\ast },  \label{fPDDNLSE}
\end{equation}%
cf. Eq. (\ref{PDDNLSE}).

Before proceeding to the numerical soliton solutions, it is instructive to examine the stability of the spatially homogeneous backgrounds, in the framework of Eq. (\ref{fPDDNLSE}). For the background $\psi_0=(A/\sqrt{2})e^{-i\theta}$ with $A\ge 0$ , the linearization of Eq. (\ref{fPDDNLSE}) with respect to the perturbation $\delta\psi \sim e^{\lambda t +ikx}$ yield the dispersion relation
\begin{equation}
	(\lambda+\gamma)^2 = h^2 - (1-2A^2+|k|^\alpha)^2 - A^2(A^2-2) , \label{eq:lambda}
\end{equation}
where $A$ satisfies the background condition $h^2=(A^2-1)^2+\gamma^2$. 
For the zero background ($A=0$), Eq.~(\ref{eq:lambda}) simplifies to
\begin{equation}
	(\lambda+\gamma)^2 = h^2 - (1+|k|^\alpha)^2.
\end{equation}
Thus, the zero solution is stable if
\begin{equation}
	h \le \sqrt{1+\gamma^2}, \label{eq:zero_stable}
\end{equation}
irrespective of the value of the Lévy index $\alpha$. This condition is the prerequisite for the existence of stable solitons that decay at infinity.

For the nonzero backgrounds, which exist at $h>\gamma$, we have $A^2 = 1\pm\sqrt{h^2-\gamma^2}$. Substituting this in Eq.~(\ref{eq:lambda}) yields the dispersion relation in the form of
\begin{equation}
	(\lambda+\gamma)^2 = 1+\gamma^2 - \left(1 \pm 2\sqrt{h^2-\gamma^2} - |k|^\alpha\right)^2. \label{eq:nonzero}
\end{equation}
For any $h>\gamma$, one can always find wavenumber $k$ such that the right-hand side of Eq.~(\ref{eq:nonzero}) exceeds $\gamma^2$, producing Re$(\lambda)>0$, hence, all the nonzero-background solutions are linearly unstable. For this reason, no stable localized states can be built upon such backgrounds, restricting the study to fully localized solitons.
 
Complex stationary solutions to Eq. (\ref{fPDDNLSE}) are looked for as

\begin{equation}
\psi (x)=\psi _{r}(x)+i\psi _{i}(x),  \label{Static-ansatz}
\end{equation}%
with real functions $\psi_{\text{r} }(x)$ and $\psi _{\text{i}}(x)$
satisfying the coupled equations%
\begin{eqnarray}
&&\left[ -\left( -\frac{d^{2}}{dx^{2}}\right) ^{\alpha /2}+2\left( \psi
_{r}^{2}+\psi _{i}^{2}\right) -\left( 1+h\right) \right] \psi _{r}  \notag \\
&=&\gamma \psi _{i},  \label{StaticEq1} \\
&&\left[ -\left( -\frac{d^{2}}{dx^{2}}\right) ^{\alpha /2}+2\left( \psi
_{r}^{2}+\psi _{i}^{2}\right) -\left( 1-h\right) \right] \psi _{i}  \notag \\
&=&-\gamma \psi _{r}.  \label{StaticEq2}
\end{eqnarray}

\begin{figure*}[th]
\centering
\hspace*{0cm} \vspace{0cm} \includegraphics[width=\textwidth]{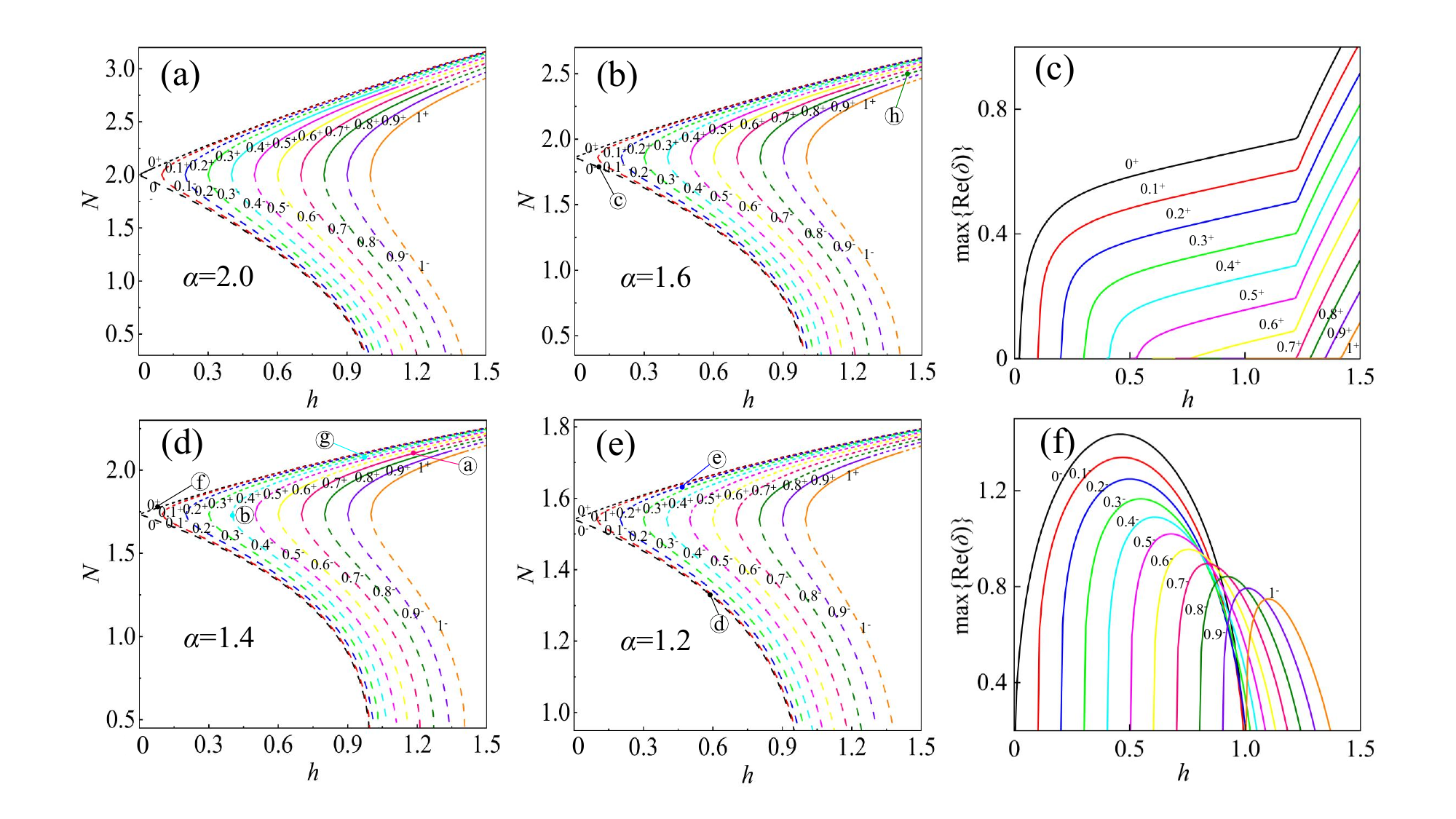}
\vspace{-0cm}
\caption{The existence and stability diagrams of the fractional quiescent
soliton solutions in the conservative ($\protect\gamma =0$) and dissipative (%
$\protect\gamma >0$) versions of the model, for different values of the loss
coefficient $\protect\gamma $ and LI values: (a) $\protect\alpha =2.0$, (b) $%
\protect\alpha =1.6$, (d) $\protect\alpha =1.4$,
and (e) $\protect\alpha =1.2$, are represented by dependences of the
soliton's norm $N$ on the parametric-gain parameter $h$. Solid lines denote
stable solitons $\psi _{+}$, while dotted and dashed lines represent
unstable solitons $\psi _{+}$ and unstable solitons $\psi _{-}$,
respectively. Labels attached to the branches mark the corresponding values of $\gamma$, with the superscripts $+$ and $-$ indicating the solitons for the $\psi _{+}$ and $\psi _{-}$ types, respectively.The evolution of solitons corresponding to the encircled labels is shown in  Fig. \ref{Fig3}. The maximum instability growth rates max {Re ($\delta$)} for $\protect\alpha =1.4$ are plotted in panels (c) and (f) for the upper and lower branches, respectively.}
\label{Fig1}
\end{figure*}

Note that, in the absence of the loss ($\gamma =0$), the non-fractional ($%
\alpha =2$) limit form of Eqs. (\ref{StaticEq1}) and (\ref{StaticEq2})
coincides with the stationary version of the integrable Manakov's system, in
which the components $\psi _{r}$ and $\psi _{i}$ have different chemical
potentials, $1\pm h$. The latter system is integrable as a system of coupled
ordinary differential equations \cite{39}, and its exact soliton solutions
are known in a cumbersome form \cite{40}.

A straightforward corollary of Eq. (\ref{fPDDNLSE}) is the evolution
equation for norm of the solution (integral power, in terms of optics), $%
N=\int_{-\infty }^{+\infty }\left\vert \psi \right\vert ^{2}dx,$ \textit{viz}%
.,%
\begin{equation}
\frac{dN}{dt}=-2\gamma N-2h\int_{-\infty }^{+\infty }\text{Im}\left( \psi
^{2}\right) dx.  \label{N}
\end{equation}%
Due to the obvious relation $\left\vert \mathrm{Im}\left( \psi ^{2}\right)
\right\vert \leq |\psi |^{2}$, Eq. (\ref{N}) implies that a stationary
soliton, with $dN/dt=0$, may exist only for
\begin{equation}
h^{2}\geq \gamma ^{2}  \label{hgamma}
\end{equation}%
(in particular, this conclusion is true for the exact solution given by Eqs.
(\ref{Exact-sols}) - (\ref{Amp})).

To produce numerical soliton solutions of the fractional equations (\ref%
{StaticEq1}) and (\ref{StaticEq2}), we applied the Newton-conjugate-gradient
method \cite{41,42} to Eqs. (\ref{StaticEq1}) and (\ref{StaticEq2}). The
stability of the resulting solutions was explored by means of adding small
perturbations to them, which were taken as%
\begin{equation}
\psi _{\mathrm{pert}}\left( x,t\right) =\left[ u(x)+iv(x)\right] e^{\delta
t},  \label{Perturbation}
\end{equation}%
where $u$ and $v$ are the real and imaginary parts of the small-perturbation
eigenmode, and Re$(\delta )$ is the instability growth rate. Substituting
the perturbation in Eq. (\ref{fPDDNLSE}) and performing the standard
linearization procedure leads to the eigenvalue problem for $\delta $,

\begin{equation}
\left(
\begin{array}{cc}
L_{11} & L_{12} \\
L_{21} & L_{22}%
\end{array}%
\right) \left(
\begin{array}{c}
u \\
v%
\end{array}%
\right) =\left( \delta +\gamma \right) \left(
\begin{array}{c}
u \\
v%
\end{array}%
\right) ,  \label{LSA-Eq}
\end{equation}%
where the component operators are%
\begin{eqnarray}
L_{11} &=&-4\psi _{r}\psi _{i},  \label{Operator-L11} \\
L_{12} &=&+\left( -\frac{d^{2}}{dx^{2}}\right) ^{\alpha /2}-2\left( \psi
_{r}^{2}+3\psi _{i}^{2}\right) +1-h,  \label{Operator-L12} \\
L_{21} &=&-\left( -\frac{d^{2}}{dx^{2}}\right) ^{\alpha /2}+2\left( 3\psi
_{r}^{2}+\psi _{i}^{2}\right) -1-h,  \label{Operator-L21} \\
L_{22} &=&+4\psi _{r}\psi _{i}.  \label{Operator-L22}
\end{eqnarray}%
The soliton is unstable if the solution of Eq. (\ref{LSA-Eq}) produces at
least one eigenvalue with Re$({\delta )}>0$. Equation (\ref{LSA-Eq}) was
solved by means of the Fourier collocation method \cite{42}. To this end,
the Fourier decomposition over a set of $512$ modes converts the
pseudodifferential equations into a matrix eigenvalue problem for Fourier
coefficients of eigenfunctions $u$ and $v$.

Existence and stability diagrams for the quiescent solitons are summarized
in Fig. \ref{Fig1}, where two types of the quiescent solitons $\psi _{\pm }
$ (cf. Eq. (\ref{Exact-sols})) are presented, for different fixed values of
LI $\alpha $ and loss constant $\gamma $ (including $\gamma =0$) by upper
and down branches of the dependence of the soliton's norm of the drive's
strength $h$. It is seen that, in agreement with Eq. (\ref{hgamma}), both
branches emerge through a bifurcation (of the saddle-node type) precisely at
$h=\gamma $, and exist at $h>\gamma $. Figures. \ref{Fig1}(c) and (f) show the maximum instability growth rate of $\psi _{+}$ and $\psi _{-}$, respectively, as a function of the parameter $h$ for $\alpha = 1.4$. The numerical stability analysis demonstrates that the upper-branch modes $\psi _{+}$ are stable between the bifurcation point $h=\gamma $ and a certain critical point $h=h_{c}$, above which they lose their stability, while all soliton solutions $\psi _{-}$ belonging to the lower branches are completely unstable.
The critical values $h_{c}$ for different values of $\gamma $ and LI $\alpha
$ are collected in Table \ref{table1}, in which \textquotedblleft N" implies
the absence of stable solitons.

\begin{table}[tbph]
\centering
%\vspace{0.1cm} \hspace*{1.5cm}
\begin{tabular}{c cccccc cccccc cccccc cccccc cccccc }

\hline
\diagbox{$\gamma$}{$h_c$}{$\alpha$} &  &  & $2.0$ &  &  &  &  &  & $1.8$ &
&  &  &  &  & $1.6$ &  &  &  &  &  & $1.4$ &  &  &  &  &  & $1.2$ &  &  &
\\ \hline
$0$ &  &  & 0.06 &  &  &  &  &  & 0.05 &  &  &  &  &  & N &  &  &  &  &  & N
&  &  &  &  &  & N &  &  &  \\ 
$0.1$ &  &  & 0.12 &  &  &  &  &  & 0.11 &  &  &  &  &  & N &  &  &  &  &  &
N &  &  &  &  &  & N &  &  &  \\ 
$0.2$ &  &  & 0.22 &  &  &  &  &  & 0.21 &  &  &  &  &  & N &  &  &  &  &  &
N &  &  &  &  &  & N &  &  &  \\ 
$0.3$ &  &  & 0.38 &  &  &  &  &  & 0.33 &  &  &  &  &  & 0.31 &  &  &  &  &
& N &  &  &  &  &  & N &  &  &  \\ 
$0.4$ &  &  & 1.07 &  &  &  &  &  & 0.65 &  &  &  &  &  & 0.43 &  &  &  &  &
& N &  &  &  &  &  & N &  &  &  \\ 
$0.5$ &  &  & 1.11 &  &  &  &  &  & 1.11 &  &  &  &  &  & 0.82 &  &  &  &  &
& 0.52 &  &  &  &  &  & N &  &  &  \\ 
$0.6$ &  &  & 1.16 &  &  &  &  &  & 1.16 &  &  &  &  &  & 1.16 &  &  &  &  &
& 0.76 &  &  &  &  &  & N &  &  &  \\ 
$0.7$ &  &  & 1.22 &  &  &  &  &  & 1.22 &  &  &  &  &  & 1.22 &  &  &  &  &
& 1.22 &  &  &  &  &  & 0.72 &  &  &  \\ 
$0.8$ &  &  & 1.28 &  &  &  &  &  & 1.28 &  &  &  &  &  & 1.28 &  &  &  &  &
& 1.28 &  &  &  &  &  & 0.93 &  &  &  \\ 
$0.9$ &  &  & 1.34 &  &  &  &  &  & 1.34 &  &  &  &  &  & 1.34 &  &  &  &  &
& 1.34 &  &  &  &  &  & 1.27 &  &  &  \\ 
$1$ &  &  & 1.41 &  &  &  &  &  & 1.41 &  &  &  &  &  & 1.41 &  &  &  &  &
& 1.41 &  &  &  &  &  & 1.41 &  &  &  \\ \hline 
\end{tabular}
%\vspace{0.1cm}
\caption{Critical values $h_{c}$ of the parametric gain, for different
values of the loss parameter $\protect\gamma $ and LI $\protect\alpha $,
above which the numerically found solitons $\psi _{+}$, belonging to upper
branches in Fig. \protect\ref{Fig1}, lose their stability. \textquotedblleft
N" indicates the absence of a stability region.}
\label{table1}
\end{table}

\begin{figure}[h]
\centering
\hspace*{-0.2cm} \vspace{-0cm} \includegraphics[width=9cm]{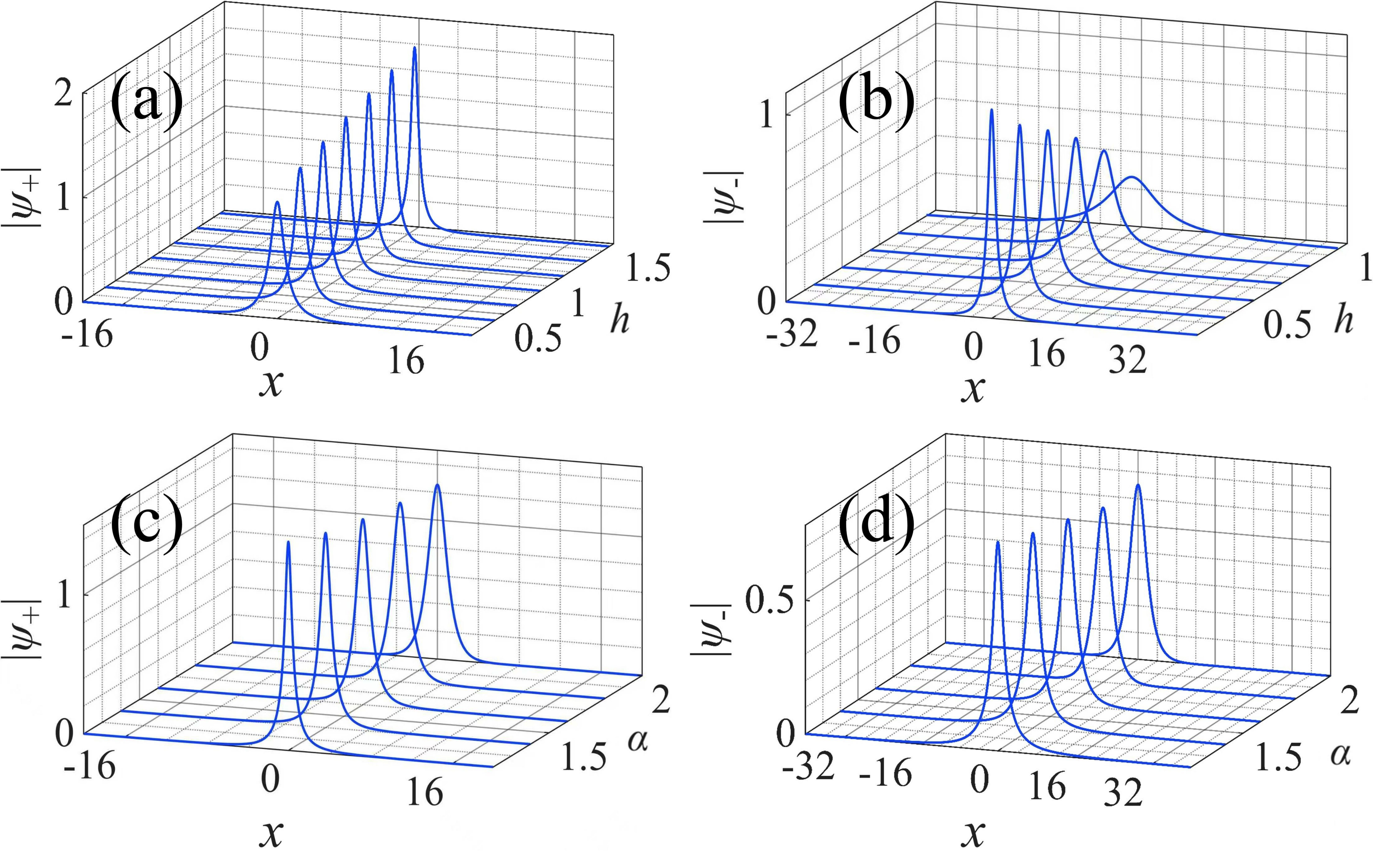} \vspace{%
		-0cm}
\caption{Profiles of solitons versus the drive's strength parameter $h$ and
the LI $\protect\alpha $. (a) and (b): The variation of $\psi _{+}$ and $%
\psi _{-}$, respectively, with $h$ at $\protect\alpha =1.4,\protect\gamma %
=0.2$. (c) and (d: The variation of $\psi _{+}$ and $\psi _{-}$, respectivel
with $\protect\alpha $ at $h=0.7,\protect\gamma =0.4$.}
\label{Fig2}
\end{figure}

Structural characteristics of the quiescent solitons exhibit distinct
dependencies on key parameters. Fig. \ref{Fig2} shows the variation of the
soliton profiles following the variation of the parameters: the amplitude of
the upper-branch $\psi _{+}$ soliton increases with the increase of the
drive's strength $h$, while the amplitude of the lower-branch $\psi _{-}$
soliton decreases, and this soliton gradually broadens as $h$ increases. As
concerns LI $\alpha $, the widths of both $\psi _{+}$ and $\psi _{-}$
solitons broaden with the increase of $\alpha $, the variation of $\psi _{+}$
being more obvious. Overall, $\alpha $ has no significant effect on the
solitons's amplitude.

\begin{figure*}[th]
\centering
\hspace*{-0cm} \vspace{-0cm} \includegraphics[width=\textwidth]{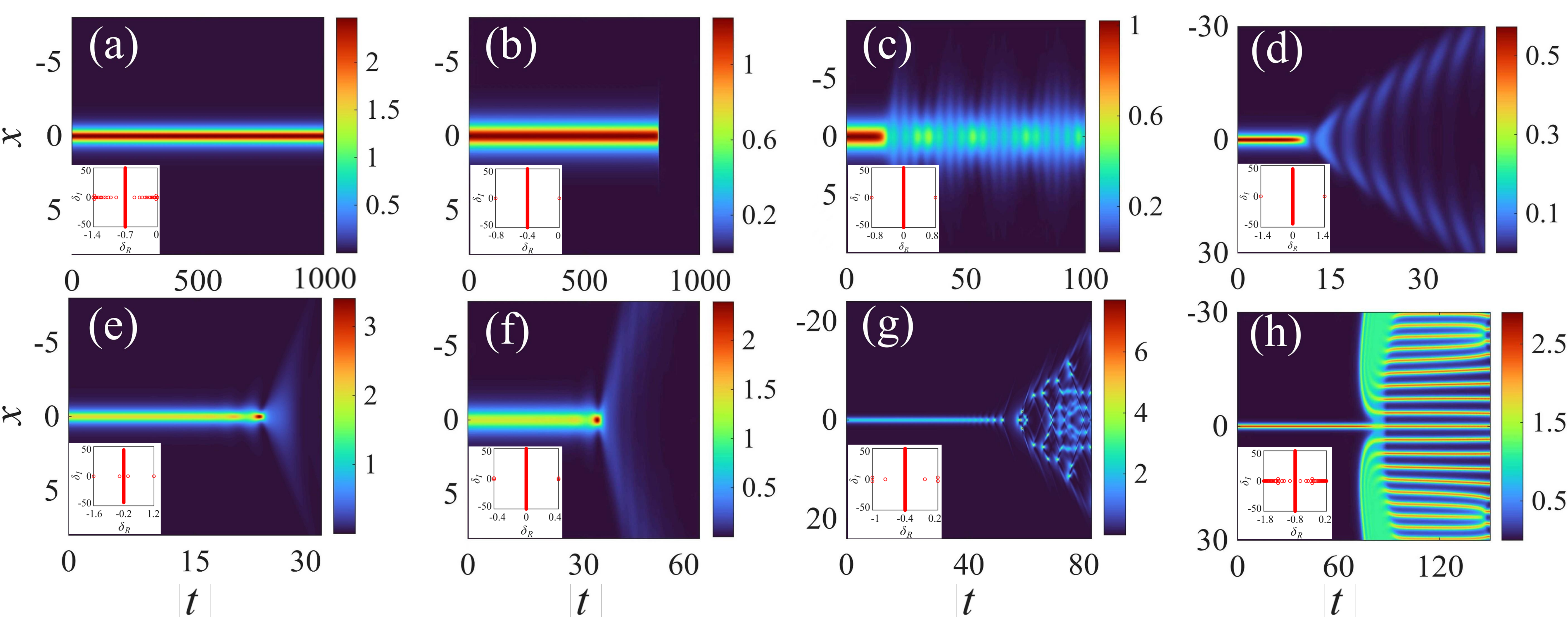}
\vspace{-0cm}
\caption{The evolution of the originally quiescent ($V=0$) solitons, corresponding to labels in Fig. \ref{Fig1}. The inset in the lower left corner of each subplot displays the corresponding linear-stability spectrum obtained by from the numerical solution of Eq. (\ref{LSA-Eq}).
 (a) The $\psi _{+}$ type with $\protect\alpha =1.4, \protect\gamma =0.7,
 h=1.2$. (b) The soliton at the critical point $(\protect\gamma =h=0.4)$ with $\protect%
\alpha =1.4$. (c) The $\psi _{-}$ type with $%
\protect\alpha =1.6, \protect\gamma =0, h=0.1$. (d) The $\psi
_{-}$ type with $\protect\alpha =1.2, \protect\gamma =0, h=0.6$%
. (e) The $\psi _{+}$ type with $\protect\alpha =1.2, \protect\gamma %
=0.2, h=0.5$. (f) The $\psi _{+}$ type with $\protect\alpha =1.4,
\protect\gamma =0, h=0.1$. (g) The $\psi _{+}$ type with $\protect%
\alpha =1.4, \protect\gamma =0.4, h=1$. (h) The $\psi _{+}$
type with $\protect\alpha =1.6, \protect\gamma =0.8, h=1.45$.}
\label{Fig3}
\end{figure*}

In addition to the static profiles, governed by parameters $h$ and $\alpha $%
, the dynamical behavior of the originally quiescent solitons in the course
of the propagation is crucially important for applications. Fig. \ref{Fig3}
displays the evolution of several typical solitons. As shown in Fig. \ref%
{Fig3}(a), the stable soliton of the $\psi _{+}$ type keeps its shape
undisturbed in the course of the long evolution. Fig. \ref{Fig3}(b) shows that the soliton at the critical point $(\gamma =h)$ suddenly dissipates after stable propagation over a relatively long distance.
 In Fig. \ref{Fig3}(c), the
unstable soliton of the $\psi _{-}$ type exhibits breathing oscillations
during the evolution. In Fig. \ref{Fig3}(d), another species of the unstable
$\psi _{-}$ soliton undergoes diffusion during the evolution. Further, it is
seen in Figs. \ref{Fig3}(e) and (f) that, under the action of a small
drive's strength, the unstable soliton of the $\psi _{+}$ type keeps its
integrity during a relatively short evolution stage, followed by diffractive
expansion. As seen in Fig. \ref{Fig3}(g), under the action of a large
drive's strength, the unstable soliton of the $\psi _{+}$type eventually
breaks up. Lastly, Fig. \ref{Fig3}(h) shows that, under the action of a
large drive's strength, the unstable $\psi _{+}$ soliton of another type
splits into multiple fragments. It is worthy to note that the fragmentation observed in Fig. \ref{Fig3}(h) is related to the modulational instability (MI) of the zero background. For the parameters of this panel, the condition $h > \sqrt{1+\gamma^2}$ holds (cf. Eq.~(\ref{eq:zero_stable})), meaning that the zero background is unstable against small perturbations, which seed the breakup of the soliton, built on this background, into multiple fragments. Thus, the dynamics displayed in Fig. \ref{Fig3}(h) results from the combined action of the soliton's internal instability and the MI of the zero background.

\section{Traveling solitons in the conservative medium ($\protect\gamma =0$)}

\begin{figure*}[th]
\centering
\hspace*{-0cm} \vspace{-0cm} \includegraphics[width=\textwidth]{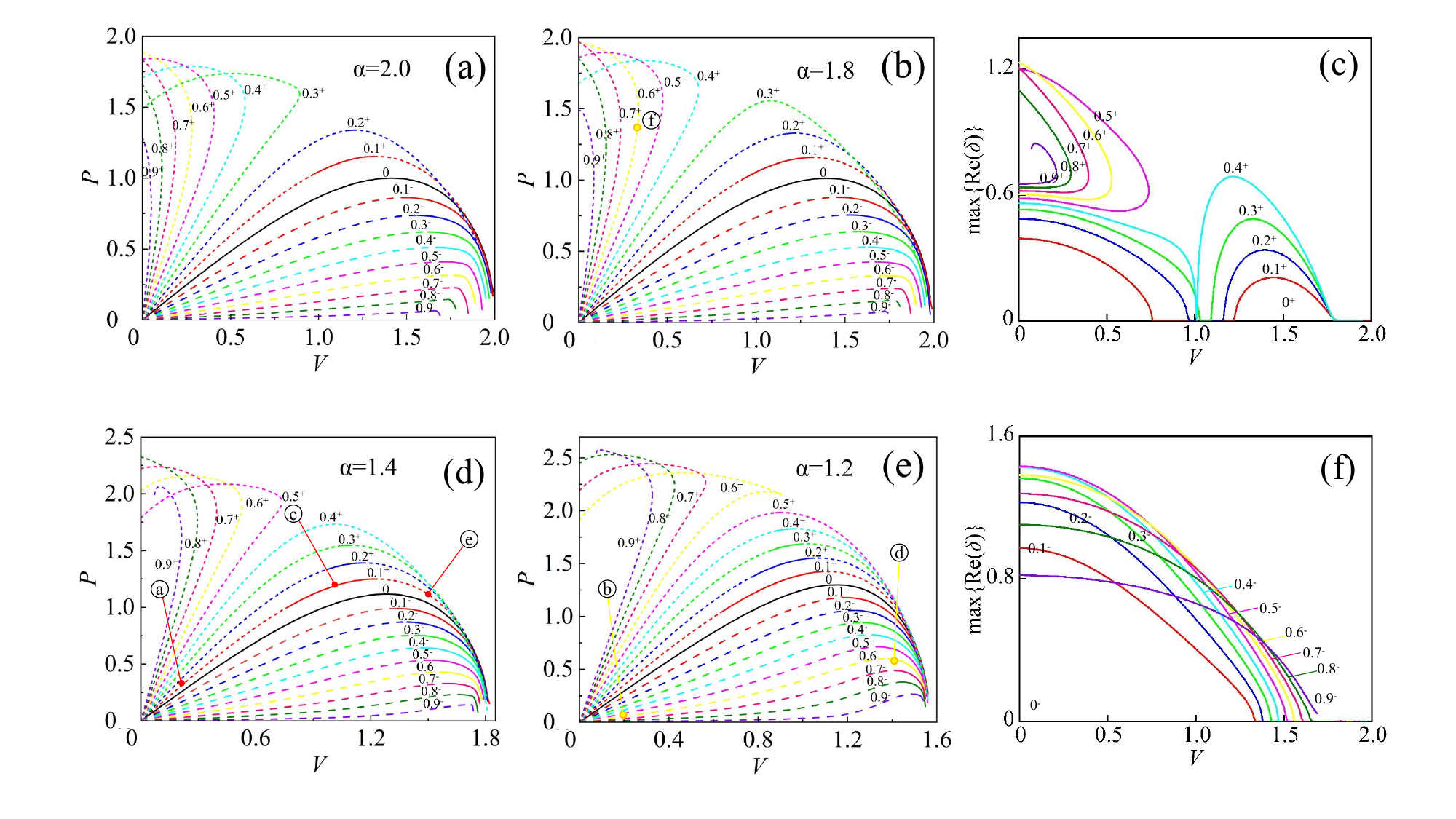}
\vspace{		-0cm}
\caption{The existence and stability diagrams for traveling solitons in the
lossless system ($\protect\gamma =0$), for different LI values(a) $\protect%
\alpha =2.0$, (b) $\protect\alpha =1.8$, (d) $%
\protect\alpha =1.4$, and (e) $\protect\alpha =1.2$, are reprsented by
dependences of the soliton's momentum $P$, defined as per Eq. (\protect\ref%
{eq:momentum}), on velocity $V$. The solitons of the $\psi _{+}$ and $\psi
_{-}$ types are located above and below the black lines $h=0$. The solid and
dashed lines denote stable and unstable solitons, respectively. Labels on the curves mark the corresponding values of $h$, with the superscripts $+$ and $-$ indicating the solitons of the $\psi _{+}$ and $\psi _{-}$ types, respectively. The evolution of traveling solitons corresponding to the encircled labels is shown in  Fig. \ref{Fig6}. The maximum instability growth rates max {Re ($\delta$)} for $\protect\alpha =1.4$ are plotted in panels (c) and (f) for the solitons of the $\psi _{+}$ and $\psi _{-}$ types, respectively.}
\label{Fig4}
\end{figure*}

The existence of traveling solitons is a nontrivial issue even in the
lossless model ($\gamma =0$), because both the parametric drive and
fractional diffraction destroy the Galilean invariance of NLSE.
Several remarks are in order regarding this restriction to $\gamma = 0$. Although the parametric drive alone already violates Galilean invariance, the inclusion of damping ($\gamma > 0$) does not, in principle, preclude the existence of traveling waves. In a dissipative system, however, a uniformly translating solitary wave with constant shape and velocity would require an exact balance between parametric gain and damping. Such states have indeed been found in the standard PDDNLSE with second-order dispersion, but they typically emerge as multi-soliton complexes moving together at a common velocity rather than as isolated pulses \cite{20,20a}. Moreover, in the fractional model, the momentum of any localized solution satisfies $dP/dt = -2\gamma P$, implying that a single soliton inherently decelerates unless it forms part of a composite structure or is subject to additional external driving. We focus here on the lossless case $\gamma = 0$, which provides a clean benchmark for isolating the effects of fractional diffraction and parametric drive on the existence, stability, and interactions of moving solitons.

In the co-moving reference frame, the corresponding fractional NLSE for the
soliton with velocity $V$ takes the form of Eq. (\ref{fmPDDNLSE}) with $%
\gamma =0$:%
\begin{equation}
i\frac{\partial \psi }{\partial t}-iV\frac{\partial \psi }{\partial \xi }%
-\psi -\left( -\frac{\partial ^{2}}{\partial \xi ^{2}}\right) ^{\alpha
/2}\psi +2|\psi |^{2}\psi =h\psi ^{\ast }  \label{fmPDNLSE}
\end{equation}%
(recall $\omega =-1$ is fixed by scaling, and $\xi =x-Vt$ is the co-moving
coordinate). The solitons moving along coordinate $x$ can be looked for as
static solutions of Eq. (\ref{fmPDNLSE}). The linear-stability analysis of
these solitons leads to the eigenvalue problem which can be written in the
general form of Eq. (\ref{LSA-Eq}), with $\gamma =0$ and component operators

\begin{eqnarray}
L_{11} &=&-4\psi _{r}\psi _{i}+V\frac{d}{d\xi },  \label{VOperator-L11} \\
L_{12} &=&+\left( -\frac{d^{2}}{d\xi ^{2}}\right) ^{\alpha /2}-2\left( \psi
_{r}^{2}+3\psi _{i}^{2}\right) +1-h,  \label{VOperator-L12} \\
L_{21} &=&-\left( -\frac{d^{2}}{d\xi ^{2}}\right) ^{\alpha /2}+2\left( 3\psi
_{r}^{2}+\psi _{i}^{2}\right) -1-h,  \label{VOperator-L21} \\
L_{22} &=&+4\psi _{r}\psi _{i}+V\frac{d}{d\xi },  \label{VOperator-L22}
\end{eqnarray}%
cf, Eqs. (\ref{Operator-L11})-(\ref{Operator-L22}).

The existence and stability of the moving solitons, as obtained from the
numerical solution of the above equations, is presented in Fig. \ref{Fig4}
by the dependence of the soliton's momentum,
\begin{equation}
P=\frac{i}{2}\int_{-\infty }^{+\infty }\left( \frac{d\psi ^{\ast }}{d\xi }%
\psi -\frac{d\psi }{d\xi }\psi ^{\ast }\right) d\xi ,  \label{eq:momentum}
\end{equation}%
on velocity $V$. In the lossless model, $P$ is a conserved quantity, along
with the above-mentioned norm $N$.

 In Fig.~\ref{Fig4}, branches located above and below the bold black line (representing the fully stable soliton family corresponding to $h = 0$) represent the extension of the solitons of the $\psi_+$ and $\psi_-$ types, respectively, in terms of Fig.~\ref{Fig1}. Solid and dotted lines
denote stable and unstable solutions, respectively. Note that many stability
boundaries exactly coincide with the maxima points, at which $dP/dV=0$;
indeed, it is well known that such points are borders between stable and
unstable traveling solitons \cite{19,20,21}. Naturally, the decrease of LI $%
\alpha $, which makes the fractional-diffraction operator more nonlocal and
thus tends to delocalize stationary states, leads to shrinkage of the
interval of velocities in which the moving solitons exist.

Similar to the existence and stability chart for the quiescent solitons
displayed in Fig. \ref{Fig1}, stability boundaries for the moving solitons
of the $\psi _{+}$ type terminate at a point where the stability interval
shrinks to zero. A
noteworthy feature of Fig. \ref{Fig4}(c) is that, when $h$
is relatively small, a nonzero velocity is necessary for the stability of
the moving solitons. At larger values of $h$, the solitons are completely
unstable, similar to the fact that the quiescent solitons of the $\psi _{+}$
type are unstable at $h>h_{c}$. Note that the decrease in the values of LI $%
\alpha $ makes the stability velocity interval for the $\psi _{+}$ solitons
somewhat \emph{larger}, while in Fig. \ref{Fig1} the decrease of $\alpha $
leads to shrinkage of the stability interval, in terms of \ the parametric
gain $h$.

A noteworthy conclusion demonstrated by Fig. \ref{Fig1} is that the solitons
of the $\psi _{-}$ type, which are completely unstable in their quiescent
form, have a finite stability domain in the high-velocity
region(Fig. \ref{Fig4}(f)). When the solitons of this type are unstable, they subject to an
oscillatory instability. As $%
\alpha $ decreases, the latter stability interval shrinks.

The findings for the stability intervals of the moving solitons of the $\psi
_{+}$ and $\psi _{-}$ types at different values of LI $\alpha $ are
summarized in Tables \ref{table2} and \ref{table3}, respectively. Note that
the moving $\psi _{-}$ solitons, which are completely unstable in their
quiescent form, have a conspicuously larger stability range than their
counterparts of the $\psi _{+}$ type.

\begin{table}[tbph]
\centering
\vspace{0.1cm}
\begin{tabular}{cccccc}
\hline
\diagbox{$h$}{$V$}{$\alpha$} & $2.0$ & $1.8$ & $1.6$ & $1.4$ & $1.2$ \\
\hline
$0.1$ & [0.98,1.32] & [0.92,1.31] & [0.85,1.28] & [0.76,1.22] & [0.64,1.08]
\\ 
$0.2$ & [1.16,1.20] & [1.15,1.21] & [1.09,1.21] & [0.97,1.16] & [0.78,1.04]
\\ 
$0.3$ & N & N & [1.10,1.11] & [1.03,1.09] & [0.85,0.99] \\ 
$0.4$ & N & N & N & N & [0.88,0.96] \\ 
$0.5$ & N & N & N & N & [0.88,0.90] \\ 
$0.6$ & N & N & N & N & N \\ 
$0.7$ & N & N & N & N & N \\ 
$0.8$ & N & N & N & N & N \\ 
$0.9$ & N & N & N & N & N \\ \hline
\end{tabular}
\vspace{0.1cm}
\caption{The velocity intervals populated by \emph{stable moving solitons}
of the $\psi _{+}$ type, for different values of LI $\protect\alpha $ and
parametric-drive's strengths $h$, in the lossless system ($\protect\gamma =0$%
) \ No stable solitons exist in cases marked by \textquotedblleft N".}
\label{table2}
\end{table}

\begin{table}[tbph]
\centering
\vspace{0.1cm}
\begin{tabular}{cccccc}
\hline
\diagbox{$h$}{$V$}{$\alpha$} & $2.0$ & $1.8$ & $1.6$ & $1.4$ & $1.2$ \\
\hline
$0.1$ & [1.49,1.99] & [1.47,1.99] & [1.42,1.93] & [1.34,1.82] & [1.20,1.56]
\\ 
$0.2$ & [1.56,1.98] & [1.53,1.98] & [1.47,1.93] & [1.39,1.81] & [1.24,1.56]
\\ 
$0.3$ & [1.62,1.97] & [1.58,1.96] & [1.52,1.92] & [1.43,1.81] & [1.27,1.56]
\\ 
$0.4$ & [1.67,1.95] & [1.63,1.95] & [1.57,1.91] & [1.47,1.80] & [1.31,1.56]
\\ 
$0.5$ & [1.72,1.93] & [1.67,1.93] & [1.61,1.90] & [1.52,1.79] & [1.34,1.56]
\\ 
$0.6$ & [1.77,1.89] & [1.72,1.89] & [1.66,1.88] & [1.56,1.78] & [1.38,1.56]
\\ 
$0.7$ & [1.77,1.85] & [1.78,1.85] & [1.70,1.85] & [1.61,1.77] & [1.41,1.56]
\\ 
$0.8$ & [1.74,1.78] & N & N & [1.66,1.76] & [1.45,1.54] \\ 
$0.9$ & [1.67,1.69] & N & N & N & [1.49,1.54] \\ \hline
\end{tabular}
\vspace{0.1cm}
\caption{The same as in Table \protect\ref{table2}, but for the moving
solitons of the $\psi _{-}$ type.}
\label{table3}
\end{table}

\begin{figure*}[th]
\centering
\hspace*{-0cm} \vspace{-0cm} \includegraphics[width=\textwidth]{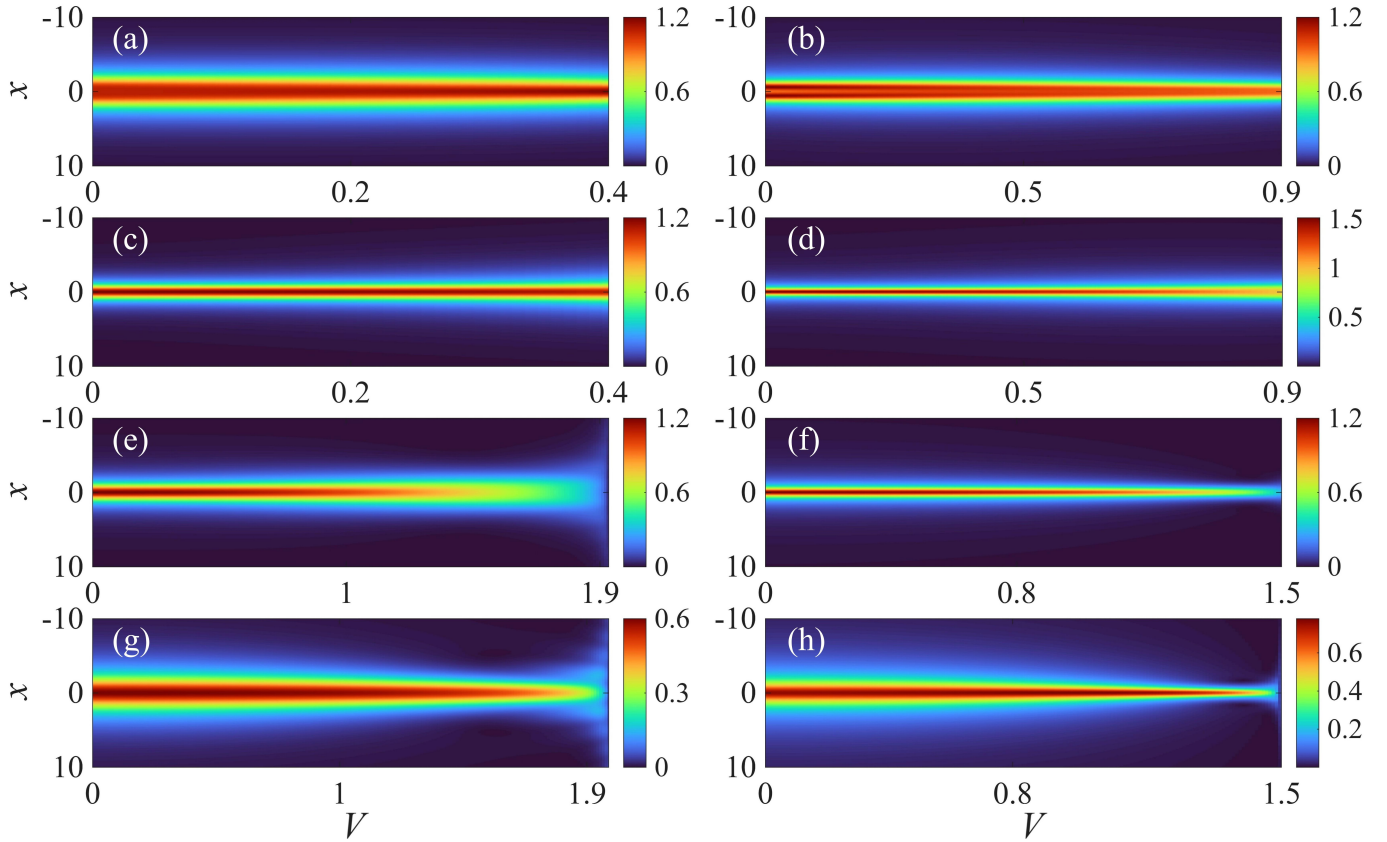}
\vspace{		-0cm}
\caption{The variation of profiles of the traveling solitons, following the
variation of velocity $V$. The figure is arranged in four rows and two
columns: the four left panels correspond to $\protect\alpha =1.6$, and the
four right ones to $\protect\alpha =1.2$. The first and second rows
represent the upper and lower branches of $\psi _{+}$, in the case when they
coexist; the third row represents the single branch of $\psi _{+}$, and the
fourth row represents the single branch of $\psi _{-}$. Parameter $h$ is  $%
0.6$ for panels (a), (b), (c), (d), (g), (h), and $0.2$ for panels (e), (f).}
\label{Fig5}
\end{figure*}

Profiles of traveling solitons also vary with the velocity $V$ and LI $%
\alpha $, as shown in Fig. \ref{Fig5}. It is arranged in four rows and two
columns: the four panels on the left and right correspond to $\alpha =1.6$
and the $\alpha =1.2$, respectively; the first and second rows represent the
upper and lower branches of $\psi _{+}$, in the case when they coexist; the
third row represents the single branch of $\psi _{+}$, and the fourth row
represents the single branch of $\psi _{-}$. For
the double-branch $\psi _{+}$solitons corresponding to large values of the
drive's strength $h$, the upper-branch solitons generally have wider
profiles, while the lower-branch ones are narrower, and their amplitudes
shows no obvious variation with $V$. The amplitude of the single-branch
traveling solitons gradually diminishes as $V$ increases. The comparison
between the left and right panels reveals that larger $\alpha $ leads to
wider soliton profiles: for a large $\alpha $ , the single-branch $\psi _{+}$
soliton (Fig. \ref{Fig5}(e)) exhibits an expanded profile and weaker
localization at larger $V$, while the single-branch $\psi _{-}$soliton (Fig. %
\ref{Fig5}(g)) forms a diffractive multi-peak structure at large $V$.
Smaller $\alpha $ results in narrower soliton profiles and higher
localization; in particular, the upper-branch $\psi _{+}$ soliton,
corresponding to smaller $\alpha $, shows a double-peak profile at low
velocity $V$, as seen in Fig. \ref{Fig5}(b).

\begin{figure}[h]
\centering
\hspace*{-0.2cm} \vspace{-0cm} \includegraphics[width=8.5cm]{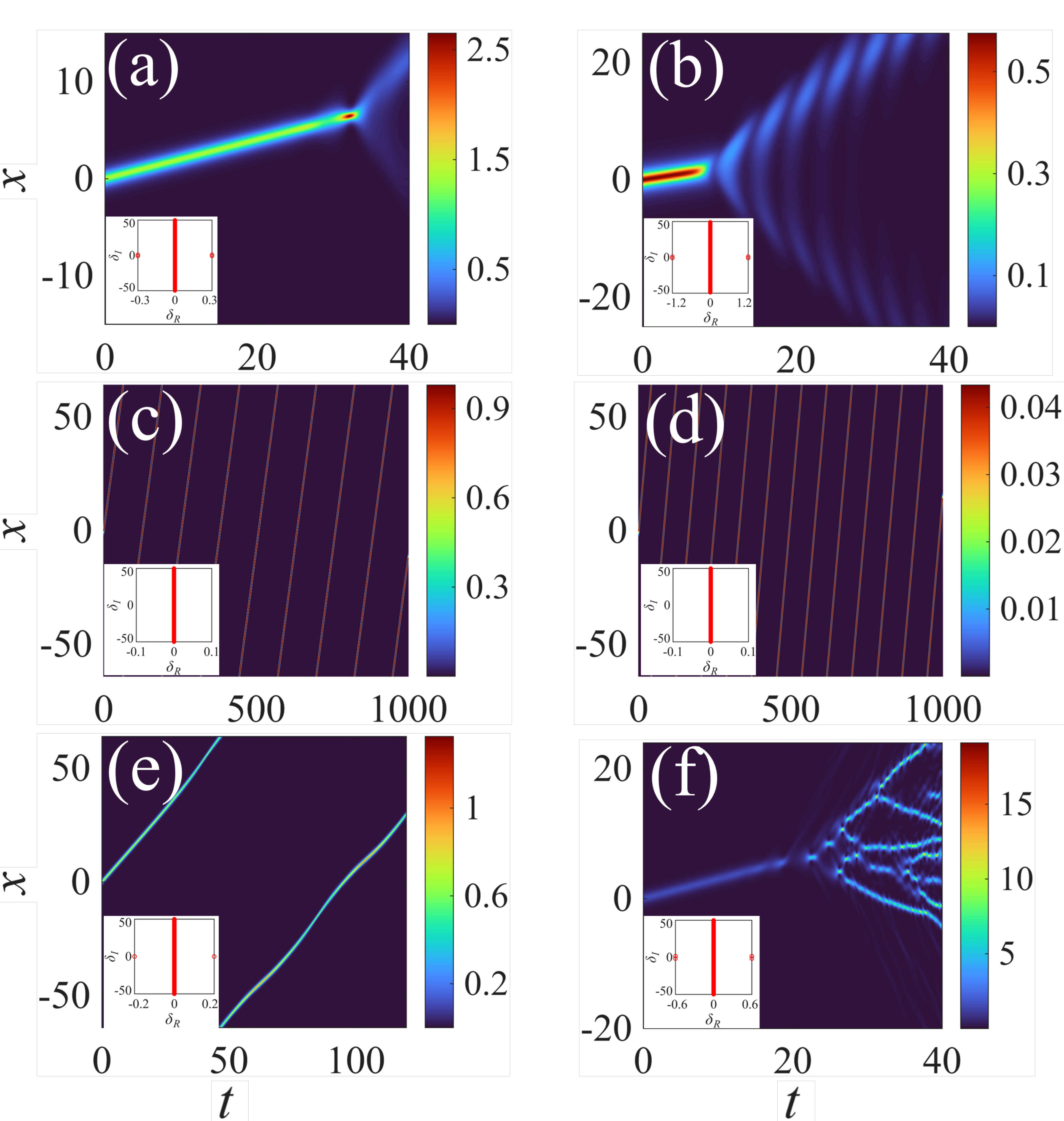}
\vspace{		-0cm}
\caption{The evolution of traveling ($V\neq0$) solitons, corresponding to labels in Fig. \ref{Fig4}. Insets in the lower left corner of each panel represent the corresponding linear-stability spectra. Panels (a),
(c) and (e) show single-branch $\psi _{+}$ solitons with $\protect\alpha=1.4,
 h=0.1$; panels (b) and (d) show single-branch $\psi _{-}$ solitons
with $\protect\alpha=1.2, h=0.6$; panel (f) shows double-branch $\psi
_{+}$ solitons with $\protect\alpha=1.8, h=0.6$. The corresponding
velocity for each panel are: (a) $V=0.2$, (b) $V=0.2$, (c) $V=1$, (d) $%
V=1.55$, (e) $V=1.5$, (f) $V=0.3$. }
\label{Fig6}
\end{figure}

It is relevant to explore the evolution of the moving solitons in the course
of the propagation. The evolution of several typical traveling solitons is
displayed in Fig. \ref{Fig6}. The traveling soliton of the single-branch
type $\psi _{+}$ is unstable at low velocities $V$, featuring undisturbed
propagation at the initial stage, which is followed by diffraction (Fig. \ref%
{Fig6}(a)). Medium velocities correspond to a stability region, allowing
long-distance stable propagation (Fig. \ref{Fig6}(c)). The propagation
becomes weakly unstable at high velocities (Fig. \ref{Fig6}(e)).  The traveling soliton of
the single-branch $\psi _{-}$ type is unstable at low velocities, exhibiting
oscillatory diffusion (Fig. \ref{Fig6}(b)), while it propagates stably at
high velocities (Fig. \ref{Fig6}(d)). The traveling soliton of the
double-branch $\psi _{+}$ breaks up in the course of the propagation (Fig. %
\ref{Fig6}(f)).

\section{Interactions between traveling solitons in the lossless medium ($%
\protect\gamma =0$)}

Interactions between solitons is a core issue in the study of nonlinear wave
systems. To simulate the interactions between moving solitons, we took the
input $\psi (x,t=0)=\psi _{1}(x-x_{1})+\psi _{2}(x-x_{2})$ for different
cases: for the head-on collision scenario, the solitons have opposite
velocities with equal absolute values (in this case, one takes the input
with $\psi _{2}=\psi _{1}^{\ast }$); for the pursuit scenario, the solitons
move in the same direction, but with different velocities, the initial
position of the faster soliton being set behind the slower one, to initiate
the collision between them.

\begin{figure}[h]
\centering
\hspace*{-0.2cm} \vspace{-0cm} \includegraphics[width=9cm]{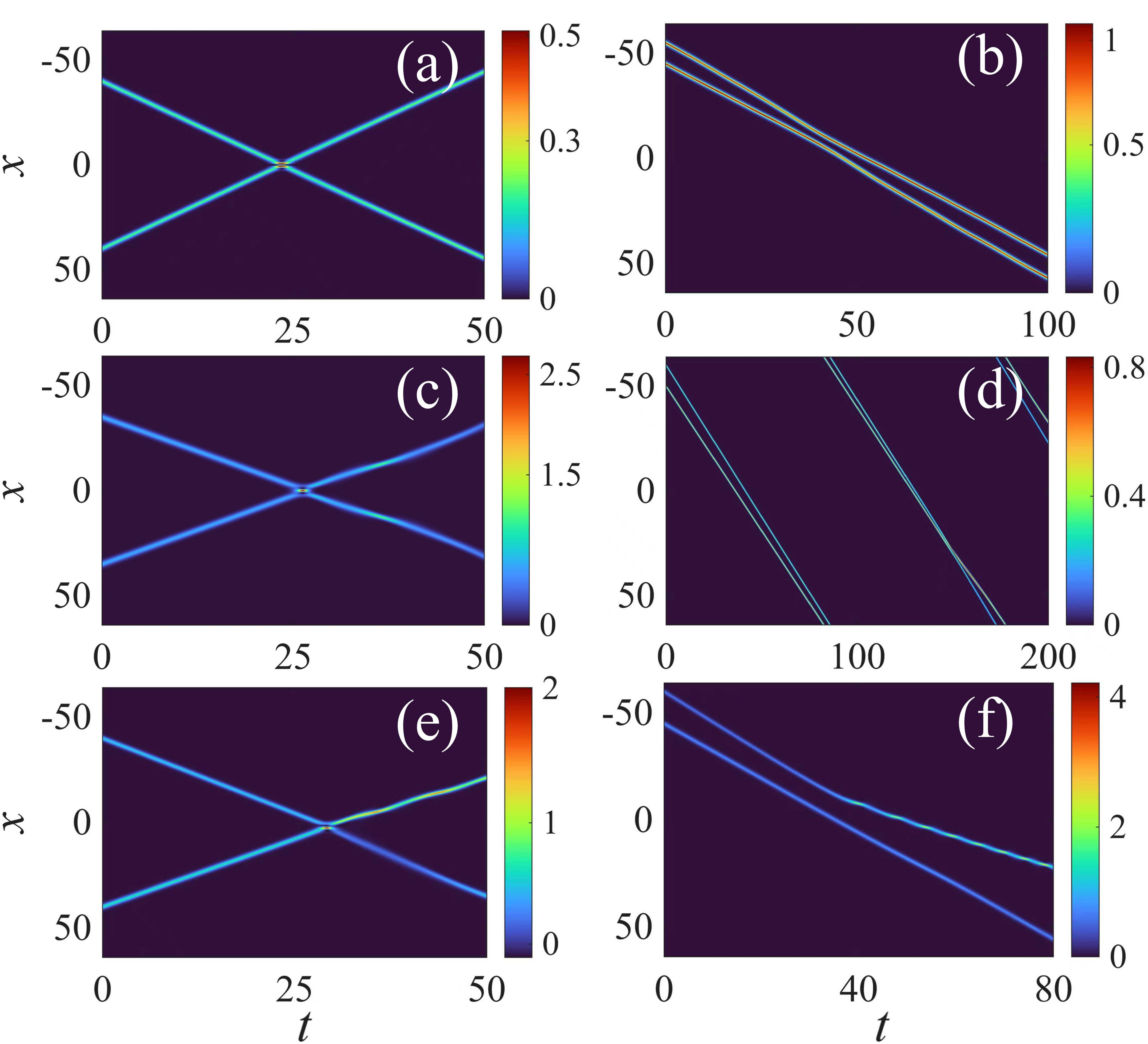} \vspace{%
		-0cm}
\caption{ The collision and pursuit interaction scenarios for pairs of
stable traveling solitons. (a) The head-on ollision between solitons of the $%
\psi _{-}$ type, with parameters $V_{1}=-V_{2}=1.7$, $x_{1}=-40$, $x_{2}=+40$%
, $\protect\alpha =1.4$. (b) The pursuit interaction between solitons of the
$\psi _{+}$ type, with parameters $V_{1}=1$, $V_{2}=0.95$, $x_{1}=-60$, $%
x_{2}=-50$, $\protect\alpha =1.4$. (c) The head-on collision between
solitons of the $\psi _{+}$ type, with parameters $V_{1}=-V_{2}=1.3$, $%
x_{1}=-30$, $x_{2}=+30$, $\protect\alpha =1.8$. (d) The pursuit interaction
between solitons of the $\psi _{-}$ type, with parameters $V_{1}=1.45$, $%
V_{2}=1.38$, $x_{1}=-60$, $x_{2}=-55$, $\protect\alpha =1.2$; (e) Collision
between $\psi _{-}$ and $\psi _{+}$ type solitons, with parameters $%
V_{1}=1.42$, $V_{2}=-1.28$, $x_{1}=-40$, $x_{2}=+40$, $\protect\alpha =1.6$.
(f) The pursuit interaction between solitons of the $\psi _{-}$ and $\psi
_{+}$ types, with parameters $V_{1}=1.42$, $V_{2}=1.28$, $x_{1}=-60$, $%
x_{2}=-50$, $\protect\alpha =1.6$. Other parameters are $h=0.1$, $\protect%
\gamma =0$.}
\label{Fig7}
\end{figure}

Under fixed parameters ($h=0.1$, $\gamma =0$), typical results of the
interactions between the solitons are summarized as follows. Fig. \ref%
{Fig7}(a) shows the head-on collision of $\psi _{-}$ type solitons. As shown
above, these moving solitons maintain their stability only at high
velocities. A pair of $\psi _{-}$ type solitons moving in the opposite
directions collide, exhibiting a strong repulsive interaction. The solitons
are compressed by the repulsion forces, and the collision is dominated by
the momentum exchange, with negligible energy variations. After the
collision, the solitons retain their original waveforms and propagate stably
in the opposite directions, without significant perturbations. Thus, the
head-on collision is quasi-elastic in this case, primarily because the two
solitons collides at high velocities, which effectively suppresses waveform
distortion and energy dissipation. Fig. \ref{Fig7}(b) represents the
pursuit interaction of $\psi _{+}$ type solitons, the velocity of the
leading soliton ($V_{2}=0.95$) being lower than that of the trailing one ($%
V_{1}=1$). Conspicuous repulsive interaction emerges as the two solitons
approach each other, which effectively prevents overlap and fusion between
them, keeping the solitons at a certain distance. After the interaction, the
solitons maintain the integrity of their profiles, without distortion,
splitting, or diffusion.

Fig. \ref{Fig7}(c) shows the head-on collision of $\psi _{+}$ type
solitons, which are stable in the intermediate velocity range. During the
approaching stage of the collision, the nonlocal coupling effect of
fractional diffraction plays a dominant role, inducing obvious energy
aggregation and thus exhibiting an attractive interaction. After the
collision, the moving solitons feature slight perturbations, mainly
manifested as tiny fluctuations in the propagation velocities. However, no
structural damage or substantial energy dissipation occurs, making the
result quasi-elastic.

Fig. \ref{Fig7}(d) depicts the pursuit interaction of $\psi _{-}$ type
solitons. This case significantly differs from the pursuit interaction of
the $\psi _{+}$ type solitons, as no obvious repulsive interaction is
observed between the two solitons. Instead, mutual penetration occurs,
accompanied by a significant energy exchange.

Figs. \ref{Fig7}(e) and \ref{Fig7}(f), respectively, show the head-on
collision and pursuit interactions between $\psi _{-}$ and $\psi _{+}$ type
solitons. In Fig. \ref{Fig7}(e), when the\ solitons of the two different
types collide, an obvious repulsive interaction is exhibited, accompanied by
energy exchange during the collision. After the redistribution of energy,
the $\psi _{+}$ type soliton maintains its stable form, while the soliton of
the $\psi _{-}$ type shows significant perturbations. In the case of the
pursuit interaction in Fig. \ref{Fig7}(f), the $\psi _{+}$ type soliton
maintains its original trajectory, unaffected by the interaction, and
demonstrating stable long-distance propagation. In contrast, the $\psi _{-}$
type soliton is repelled during the interaction, showing obvious
oscillations in its trajectory and failing to maintain a stable form.

\section{Conclusion}

This work aims to systematically explore the existence, stability, and
interactions of solitons in the model based on the PDDNLSE (parametrically
driven damped nonlinear Schr\"{o}dinger equation) with the diffraction
operator represented by fractional Riesz derivative, whose LI (L\'{e}vy
index) $\alpha $ takes value in the relevant interval, $1<\alpha \leq 2$.
Recent experimental results \cite{25,43} suggest that the model can be
realized in optics, with the fractional diffraction or dispersion emulated
by means of the specially designed cavity \cite{50}. Two families of the
quiescent solitons ($\psi _{+}$, which is stable in a finite interval of
values of the parametric gain $h$, and completely unstable $\psi _{-}$) are
found. The stability interval of the solitons $\psi _{+}$ shrinks with the
decrease of $\alpha $, disappearing at $\alpha =1$. In the lossless system ($%
\gamma =0$), traveling solitons exist with velocity $V$ restricted by the
maximum value, which decreases with the decrease of $\alpha /h$. The moving
solitons of the $\psi _{+}$ type are stable at intermediate velocities for
small $h$, and completely unstable for larger $h$, while the moving solitons
of the $\psi _{-}$ are stable at high velocities. The decrease of $\alpha $
leads to the shrinkage of the latter high-velocity stability region.
Repulsive collisions between\ stable $\psi _{-}$ solitons at high velocities
are quasi-elastic, whereas collisions between $\psi _{+}$ solitons, which
are stable in the intermediate velocity range, are dominated by the
effective attraction, induced by the nonlocal coupling. Overall, the
fractional diffraction strongly affects properties of the solitons in the
lossy and lossless systems, providing guidance for new experiments.

As an extension of the present analysis, it is relevant to address the
existence and stability of bound states of two or several solitons,
quiescent and moving ones, cf. Ref. \cite{16}, where bound states were found
in the nonfractional model.

\begin{acknowledgments}
	This research was supported by the National National Natural Science Foundation of China (Grant Nos. 11805141, 12104353), the Fundamental Research Program of Shanxi Province (Grants No. 202303021211185), National Key Research and Development Program of China (Grant No. 2022YFA1404902), Fundamental Research Funds for the Central Universities (Grant No. QTZX25086), the Israel Science Foundation (Grant No. 1695/22), and from ANID (Chile) through FONDECYT (Grant No. 1260401).
\end{acknowledgments}

\section*{Data Availability}

The data that support the findings are available from the authors upon
reasonable request.

\end{document}